\documentstyle[eqsecnum,multicol,psfig,aps,prl,array]{revtex}
\newcommand{\bleq}{\ifpreprintsty
                   \else
                   \end{multicols}\widetext \vspace*{-3.5ex}{\tiny
                   
                \noindent\begin{tabular}[t]{c|}
                   \parbox{0.493\hsize}{~} \\ \hline \end{tabular}}
                                      \fi}
\newcommand{\eleq}{\ifpreprintsty
                   \else
                   {\tiny\hspace*{\fill}\begin{tabular}[t]{|c}\hline
                    \parbox{0.49\hsize}{~} \\
                    \end{tabular}}\vspace*{-2.5ex}\begin{multicols}{2}
                    \narrowtext
                    \fi}
\newcommand{\bcols}{\ifpreprintsty\else\begin{multicols}{2} 
        \narrowtext\fi}
\newcommand{\ecols}{\ifpreprintsty\else\end{multicols}\fi}

\draft
\widetext
\begin{document}
\title{Glass transition temperature variation, cross-linking and structure 
in network glasses: a stochastic approach} 
\author{Matthieu Micoulaut $^{\S}$ and Gerardo G. Naumis $^{\P}$}
\address{ \\
$^{\S}$ Laboratoire GCR-CNRS URA 769\\
Universit\'e Pierre et Marie Curie, Tour 22, Boite 142\\
4, Place Jussieu, 75252 Paris Cedex 05, France\\
$^{\P}$ Instituto de Fisica, Universidad Nacional Autonoma de M\'exico\\
Apartado Postal 20-364, 01000 M\'exico DF, M\'exico}
\begin{abstract}
Stochastic network description provide useful information about the link 
between glass transition temperature $T_g$ and network connectivity.
In multicomponent glasses, this
permits to distinguish homogeneous compositions (random network) from 
inhomogeneous ones (local phase separation). The stochastic origin of the
Gibbs-Di Marzio equation is predicted at low connectivity and the analytical
expression of its parameter emerges naturally from the calculation.\par
Pacs numbers: 61.20N-81.20P
\end{abstract}
\bcols
Most inorganic solids can be made amorphous by vapor deposition onto cold
substrates. However, only a very few of inorganic melts can be 
supercooled by a water or air quench to yield bulk glasses which solidify at
the glass transition temperature $T_g$. Oxides as vitreous silica ($SiO_2$)
and chalcogenides (e.g. $Ge_xSe_{1-x}$) represent some of the best-known glass
formers in nature. 
There have been numerous efforts 
to understand the nature of glass transition, and to relate
$T_g$ to some easily measurable quantities. 
Tanaka has proposed a relationship between $T_g$ and the mean coordination number \cite{Tanaka}. Gibbs and Di Marzio have developed a second-order phase transition model and obtained an empirical relationship (GDM equation) between the transition temperature and the density of cross-linking agents inserted inside a system of molecular chains \cite{Adam}. 
However, there is still no universal relationship between $T_g$ and the 
glass network connectivity, satisfied by allmost all kinds of glass formers. The first
attempt of such a quantitative description has been given only very 
recently \cite{Richard}.
We present in this Letter several important results concerning the glass transition temperature  variation as a function of connectivity in network glasses. 
\par
Mean-field estimates of $T_g$ work only at low connectivity, but they do not work at high connectivity because a stochastic description fails.\par
In the chalcogen limit (low connectivity), an adapted version of the GDM 
equation for chalcogenides \cite{Sreeram} can 
be obtained analytically from the stochastic description of the network. 
Consequently, the parameter $\beta$ appearing in \cite{Sreeram} can be computed 
in this limit from the coordination number for any glass system, with simple 
and elegant sum rules.
\par
The combination of the GDM equation in \cite{Sreeram} and the topological 
calculation of the parameter $\beta$ yields the correct trends in the $T_g$ variation up to $\bar r=2.4$, 
in agreement with Phillips' constraint theory \cite{Phillips}. Moreover, it 
suggests that usual curve-fitting with GDM equation hides the stiffness transition 
at $\bar r\simeq 2.4$, when realized over the whole concentration range 
because of the occurence of chemical ordering.
\par
The prediction of the model is parameter-free and it can be easily extended from binary to ternary, quaternary and multicomponent network forming materials. We 
should stress at this point that the model is not intended to describe the 
physics of glass transition, but rather the connectivity dependence of the 
temperature of this transition. To this end, the chalcogenide systems are of 
particular interest for the application of the model because of the minor 
role played by kinetics, compared to the larger role of connectivity in 
determining the value of the glass transition temperature \cite{Boolchand}.
For the reader's convenience, let us first sketch the main ideas of the stochastic model.
\par
If the network of a certain glass system form a random network, then one 
should be able to treat statistically with equivalent fashion different states
of structural description. For example, in a binary 
$B_xA_{1-x}$ structure the mean probability $p_A^{b}=1/2[2p_{AA}+p_{AB}]$ of finding
an atom $A$ among randomly distributed bonds $A-A$ and $A-B$ (with probability 
$p_{AA}$ and $p_{AB}$) should be equal to the probability $p_A^{a}$ of 
finding it among a random distribution of $A$ and $B$ atoms. Thus, 
$p_A^b=p_A^a$. We have excluded the possibility of a B-B bond which occurs 
only in a modifier-rich glass structure. The bond probabilities $p_{ij}$ 
(with (i,j)=(A,B)) should be proportional to the concentration $x$ and  
$(1-x)$ (or $p_A^{a}$ and 
$p_B^{a}$), a statistical weight $w_{ij}$ related to the coordination numbers 
of the atoms $A$ and $B$ ($r_A$ and $r_B$), 
and a Boltzmann factor involving both the 
glass transition temperature $T_g$ and the bond energies $E_{ij}$. The 
probabilities $p_{AA}$ and $p_{AB}$ can be written as follows:
\begin{eqnarray}
\label{1}
p_{AA}={\frac {r_A^2}{\cal Z}}(1-x)^2e^{-E_{AA}/k_BT_g}
\end{eqnarray}
\begin{eqnarray}
\label{2}
p_{AB}={\frac {2r_Ar_B}{\cal Z}}x(1-x)e^{-E_{AB}/k_BT_g}
\end{eqnarray}
${\cal Z}$ normalizes the bond probabilities. 
The statement: $1-x=p_A^a=p_A^b$ can be solved in terms of the concentration $x$,
because $p_A^b$ is constructed with it:
\begin{eqnarray}
\label{3}
x={\frac {r_Be^{-E_{AB}/k_BT_g}-r_Ae^{-E_{AA}/k_BT_g}}{2r_Be^{-E_{AB}/k_BT_g}-r_Ae^{-E_{AA}/k_BT_g}}}
\end{eqnarray}
Also, equation (\ref{3}) can be made parameter-free (i.e. without involving 
$\Delta \varepsilon=E_{AB}-E_{AA}$) by condidering the limit $x=0$ (pure chalcogen) 
where $T_g(0)=T_0$. One gets from (\ref{3}):
$\Delta \varepsilon=E_{AB}-E_{AA}=k_BT_0\ln [r_B/r_A]$
Finally and most importantly, the relationship (\ref{3}) can be cast in a 
more compact presentation by performing the derivative of (\ref{3}) with 
respect to $T_g$ in the limit $x=0$, and inserting the energy difference $\Delta \varepsilon$. 
We obtain then the following parameter-free slope equation which is the 
central result \cite{Richard} to be used in the present work:
\begin{eqnarray}
\label{4}
\biggl[{\frac {dT_g}{dx}}\biggr]_{x=0,T_g=T_0}\ =\ {\frac {T_0}{\ln \biggl[{\frac {r_B}{r_A}}\biggr]}}
\end{eqnarray}
As $\bar r$ is defined by: $\bar r=r_A(1-x)+r_Bx$, one can obtain the 
derivative of $T_g$ with respect to $\bar r$ by replacing in all equations 
(\ref{1})-(\ref{3}) $x$ by $(\bar r-r_A)/(r_B-r_A)$:
\begin{eqnarray}
\label{5}
\biggl[{\frac {dT_g}{d\bar{r}}}\biggr]
_{\bar{r}=r_A,T_g=T_0}={\frac {T_0}{(r_B-r_A)\ln \biggl[{\frac {r_B}{r_A}}\biggr]}}
\end{eqnarray}
The value of the coordination number $r_B$ can be determined in most of the
situations by the $8-N$ rule, where $N$ is the number of outer shell 
electrons of the considered atom \cite{Mott} (and of course, in the forthcoming,
$r_A=2$).
From the second part of equation (\ref{5}), it is possible to demonstrate
the stochastic origin of the heuristic GDM equation at low connectivity 
($\bar r=2$) and to show that the modified equation proposed by Varshneya
and co-workers \cite{Sreeram} is the correct expression as long as this 
equation remains linearly extrapolated.
The GDM theory of glass transition  (based on equilibrium principles) is 
intended to describe the variation of $T_g$ of long polymer chains of equal 
length, with chain stiffness produced by cross-linking agents \cite{Adam}.
The $T_g$ of the cross-linked glass is suggested to behave as: 
$T_g=T_0(1-\kappa X)^{-1}$ where $X$ is the cross-link density and $\kappa$ 
a constant. From the construction of the theory, is easily conceivable that 
it might describe chalcogenide glasses as well, because the initial selenium network is 
also made of long polymeric chains, and the modifier atoms as Ge or As 
should play the cross-linking agents. To this end, the GDM equation has been 
recently adapted with success by Varshneya and co-workers in order to describe the $T_g$ 
trends in chalcogenides \cite{Sreeram}: $T_g=T_0/(1-\beta (\bar r-2))^{-1}$ 
(which we will denote by VGDM equation). The parameter $\beta$ is obtained 
from a least-squares fitting of the experimental data. Now, if one performs 
the derivative with respect to $\bar r$ of the first-order Taylor expansion 
(linear) of the latter expression in the vicinity of $\bar r=2$, one obtains 
$\beta T_0$. This can be compared with the right-hand side of equation 
(\ref{5}) and yields an analytical expression for $\beta$:
$\beta^{-1}=(r_B-2)\ln [{\frac {r_B}{2}}]$.
In a two-component chalcogenide glass (A,B), the constant $\beta$ has a topological 
origin and can be easily computed from the coordination number of the modifier 
atom B.
\par
We shall prove that the factor appearing in the expression of $\beta$ 
has a universal character and can be extended to M-component glass 
system, yielding the value of the parameter $\beta$ for any sytem, to be 
inserted in the VGDM equation. This result will still be obtained in the 
limit where the VGDM equation emerges naturally from a stochastic description 
of the network.
\par
Before, we shall consider a glass system made of three different kinds of 
atoms (say $A$, $B$ and $C$, with respective concentration $1-x-y$, $x$ and $y$), one of them 
being the chalcogenide atom of the chain-like initial structure (when $x=0$ and $y=0$). The
coordination numbers of the involved atoms are $r_A=2$, $r_B$ and $r_C$. The average 
coordination number is $\bar{r}=r_Bx+r_Cy+2(1-x-y)$ and: 
\begin{eqnarray}
\label{6}
{\frac {d\bar{r}}{dT_g}}\ =\ (r_B-2){\frac {dx}{dT_g}}+(r_C-2){\frac {dy}{dT_g}}
\end{eqnarray}
We can still identify the derivative of the first-order Taylor expansion of 
the VGDM equation in the vicinity of $\bar{r}=2$ with the right-hand side of 
(\ref{6}),where consequently $x=0$ and $y=0$ and where the quantities 
$dx/dT_g$ and $dy/dT_g$ have the form presented in equation (\ref{4}) 
(i.e. $\ln[r_B/r_A]/T_0$ and $\ln[r_C/r_A]/T_0$). By identification, this 
leads to the analytical expression of the parameter $\beta$ in a glass made of three components: $\beta^{-1}=(r_B-2)\ln [r_B/2]+(r_C-2)\ln [r_C/2]$. The extension to multicomponent systems appears to be quite natural. $\beta$ has the same sum rules as the resistance in a parallel circuit in electrokinetics, i.e. it is the sum of the $1/\beta$ of each related two-component system AB, AC, etc. Finally, for a system made of
$M$ different kinds of atoms with coordination numbers $r_i$, we just have 
to sum up the $M-1$ contributions $(r_i-2)\ln [{\frac {r_i}{2}}]$ in order 
to obtain the theoretical value of $\beta^{-1}$:
\begin{eqnarray}
\label{7}
{\frac {1}{\beta}}\ =\ \sum_{i=1}^{M-1} (r_i-2)\ln\biggl[{\frac {r_i}{2}}\biggr]
\end{eqnarray}
In this notation, $m_M$ is the coordination number of the chain atom, equal 
to $2$. However, when applied to ternary systems, the stochastic description 
with three kinds of atoms A, B and C is able to predict the 
VGDM equation up to $\bar r\simeq 2.3$ with the correct value of $\beta$, 
satisfying (\ref{7}), starting from a system which satisfies $p_B^a=p_B^b$ and
$p_C^a=p_C^b$ \cite{Naumis}.
\par
{\em Comparison with experimental data:} In order to minimize the influence 
of the preparation techniques, we have carefully selected
data of i) glass systems prepared with the same heating/cooling rate and ii) 
glass systems with more than five different compositions \cite{check}. 
The initial value $T_0$
has been averaged over a set of data found in the literature ($T_0$ of
$v-Se$ has been taken as $316\ K$, of $v-S$ as $245\ K$). We have plotted in
figure 1 several experimental data on binary, ternary and quaternary systems
with the parameter $\beta$ computed from the involved coordination numbers $r_i$
and the VGDM equation (solid lines). The dotted lines correspond to the slope 
equation (\ref{5}) and shows the stochastic origin of the VGDM equation at 
low connectivity ($\bar r\simeq 2$). 
For other systems, we have performed a least-squares fit of the parameter 
$\beta$ from the VGDM equation, denoted as $\beta_{exp}$. Then, we have 
compared the results with the predicted parameter $\beta_{pr}$ obtained from
(\ref{7}). In the system Si-As-Ge-Te the involved coordination numbers of 
the modifier atoms are $r_i=(4,3,4)$. Thus, $\beta_{pr}=(3\ln 2+\ln 3)^{-1}=
0.31$, in excellent agreement with the fit $\beta_{exp}=0.30$ \cite{ElFouly}. 
Examples displayed in fig. 1a show also that the slope (\ref{5}) yields accurate 
trends in the variation of $T_g$ with respect to 
the average coordination number $\bar{r}$ at low connectivity.
\par
From fig. 1a, we can obviously see that the 
addition of a two-coordinated atom (as tellurium or sulphur, $r_i=2$) in 
multicomponent glass systems does not affect the value of the parameter 
$\beta$ (because of the rate $\ln[ri_/2]$). A quaternary system which 
involves a two-coordinated atom can therefore be considered as a ternary 
system. The simultaneous use of the VGDM equation and the stochastic 
prediction of
(\ref{7}) can give the value of the glass transition temperature of any 
composition, at least for $\bar{r}\leq
2.4$ which is the limit imposed by Maxwell rigidity \cite{Phillips}. We 
invite the reader to check on this basis that the $T_g$ of 
$Ga_{10}B_{10}S_{80}$ is about $325\ K$.
For greater values of $\bar{r}$, one has to take into account intermediate 
range order effects such as the existence of rings \cite{Boolchand}, 
\cite{Angell} or chemical ordering. However, the stochastic model suggests 
also that usual curve-fitting with the VGDM equation \cite{Sreeram},\cite{Saiter} does 
not permit to distinguish between floppy and rigid regions (limited by 
$\bar r=2.4$), whereas the combination of VGDM and (\ref{7}) show direct
evidence of the stiffness transition between these two regions. This is due to 
the fact that $\beta$ is evaluated from the rendom network description and it
fails around $2.4$. To illustrate this 
observation, we have plotted in fig. 1b for the P-Ge-Se compound the 
least-squares fit of the VGDM equation (dotted line) ($\beta_{exp}=0.66$, 
realized over the whole concentration range \cite{Lyda}) and the VGDM 
equation, combined with the sum rule (\ref{7}) for $\beta_{pr}=0.55$ (solid line). The
deviation occurs at $\bar r=2.43$, consistently with constraint counting
arguments \cite{Phillips}. We have also plotted in the insert of fig. 1b
the value $\beta_{exp}$ computed from the VGDM equation, as a function of
the average coordination number, in In-Ge-Se systems \cite{Saiter}.
This clearly shows the threshold around $\bar r=2.4$ between the values 
$\beta=0.55$ (predicted from the model by the random network picture) and 
$\beta=0.72$ (predicted from the model by the occurence of chemical ordering,
as we shall see below). A global fit of the VGDM equation (dotted line) 
with a unique value of $\beta$ would not have given this information.
\par
{\em Effect of chemical ordering:} 
In most of the chalcogenide systems, chemical ordering occurs for $\bar r\geq 2.4$, 
and therefore stochastic description fails, as shown in fig. 1 
However, the description is still useful, if the network can be thought as
a set of compound clusters inside a random network. 
\par
Again, let us consider a ternary glass system $A_{1-x-y}B_xC_y$ 
with the corresponding coordination numbers $r_A=2$, $r_B$ and $r_C$.
If we assume that the general tendency of the glass is to form a demixed
structure of $A$ and $B$ in stoichiometric proportions, then we can rewrite
the system as: $(B_{r_A}A_{r_B})_{x/r_A}C_yA_{1-y-(r_A+r_B)x/r_A}$ . For the
glassy matrix, this defines an effective concentration of $C$ atoms 
$y^{eff}={\frac {y}{1-x(r_A+r_B)/r_A}}$
and the average coordination number of the glassy matrix is given by:
\begin{eqnarray}
\label{8}
\bar{r}&=&r_A+{\frac {(r_C-r_A)y}{1-x{\frac {(r_A+r_B)}{r_A}}}}
\end{eqnarray}
If we proceed as before, i.e. performing the derivative with respect to $T_g$ and
looking at the limit ($x,y\rightarrow 0$) in order to identify with the 
expansion of the VGDM equation, we can see that the corresponding
parameter $\beta$ is defined as $\beta^{-1}=(r_C-r_A)\ln [{\frac {r_C}{r_A}}]$.
In other words, the parameter $\beta$ of a ternary system which displays 
chemical ordering can be computed by considering only the remaining two-component
glass. We have checked the validity of this rule on a set of
germanium incorporated chalcogenides. We have considered all
possible stoichiometric demixed structures at the tie-line composition
(e.g. in $Ge_xAs_ySe_{1-x-y}$, the possible structures are $GeSe_2$ and
$As_2Se_3$). Most of the ternary $III-IV-VI$ systems such as $Sb-Ge-X$ 
glasses (X=S, $\beta_{exp}=0.61$ \cite{ElHamalawy}; X=Se, $\beta_{exp}=0.78$ 
\cite{Fouad}; X=Te, $\beta_{exp}=0.79$ \cite{Lebaudy}) behave as a single 
binary $IV-VI$ glass (as $Ge-Se$) with parameter close to $\beta_{pr}=0.72$ 
when $\bar{r}\ >2.4$. This is explained by the presence of $Sb_2X_3$ clusters
($X=S,Se,Te$) inside the remaining random network of Se-Se and Ge-Se bonds. 
Also, data on $Sn-Ge-Se$ systems show that $\beta_{exp}=0.68$, when $SnSe_2$
clusters are assumed, close to $\beta_{pr}=0.72$ \cite{Haruvi}.
Coming back to the illustrative $In-Ge-Se$ compound (insert of fig. 1b), the 
behavior 
of $T_g$ versus $\bar{r}$ can be explained as follows. For $\bar{r}< 2.4$, the structure 
can be described by a random network of $Se-Se$, $Ge-Se$, $In-Se$ and $Ge-In$
bonds. 
For this system, the VGDM equation and (\ref{7}) describe the $T_g(\bar{r})$
behavior, with $\beta_{pr}=0.55$ is computed from (\ref{7}) with $r_i=(3,4)$. 
For $\bar{r}\ >2.4$, the network structure looses its random
character because of the occurence of $In_2Se_3$ clusters \cite{Saiter}. 
Thus, the system behaves as a pseudo-binary $Ge-Se$ system with the effective 
concentration (\ref{8}) and $\beta_{pr}=0.72$. \par
The authors gratefully acknowledge R. Kerner, R.A. Barrio and J. Ledru for
stimulating discussions. This work has been supported by the CONACyT grant
No. 25237-E.

\newpage
\begin{figure}
\begin{center}
\caption{a) Comparison of the VGDM equation with the stochastic calculation of 
$\beta$ (\ref{7}) with experimental measurements 
in binary, ternary and quaternary glasses (solid line). The slope equation 
(\ref{5}) corresponds to the dotted lines. Systems $Ge-Te-(Se)$, $r_i=(4,2)$ 
and $Ge-(Se)$, $r_i=4$ [13]: $\beta_{pr}=0.72$. System $Sb-Ge-Te-(Se)$, 
$r_i=(3,4,2)$ and $Sb-Ge-(Se)$, $r_i=(3,4)$ [4]: 
$\beta_{pr}=0.55$. System $Ge-Br-(S)$, $r_i=(4,1)$ [14]: 
$\beta_{pr}=0.48$. b) Occurence of chemical ordering in ternary selenides (data from 
[4], [16], [17]). The stochastic description 
yields for all $\beta_{pr}=0.55$, since $r_i=(3,4)$ in the three systems. 
The solid line represents the combination of this description with the VGDM 
equation. The dotted line correspond to the VGDM equation with $\beta$ fitted 
from the whole concentration range, as realized in [4]. 
For $\bar r >2.4$, the systems behave as a binary glass with $r_i=4$ and 
$\beta_{pr}\simeq 0.72$. The insert shows $\beta_{exp}$ versus the average 
coordination number $\bar r$ for the $In-Ge-Se$ compound. All data sets have
been displaced by $100\ K$ for a clearer presentation.}
\end{center}
\end{figure}

\ecols


\begin{thebibliography}{100}
\bibitem{check} We have checked that the influence of the heating rate on 
the value of $\beta$
could be neglected at the heating rates which were used in the preparation of the
glasses we have investigated. To do this, we have used both the Kissinger's 
formula \cite{Kissinger}. 
$\ln [{\frac {T_g^2}{Q}}]+const.={\frac {E}{RT_g}}$,
which is very well adapted for the description of 
chalcogenide glasses, and the VGDM equation.
$Q$ is the heating rate, $E$ the activation energy for glass transition
and $R$ the gas constant. Inserting typical values of $\bar{r}$, $E$ and $T_0$
shows that $\beta$ depends weakly on $Q$ ($\Delta\beta/\beta\leq 5\%$).
This verification has been also realized on systems for which the heating
rate in DSC calorimetry was reported. 
For example, the parameter $\beta$ of the glass $As-Sb-Se$ lies in the range 
[1.14, 1.17] (computed from a least-squares fit) for 
cooling rates between $Q=0.62\ K.min^{-1}$ and $40\ K.min^{-1}$, which 
corresponds to $\Delta \beta/\beta=0.013$ \cite{Mahadevan}. In $Ge-Sb-Se$ 
glasses, $\beta$
varies as $\Delta \beta/\beta =0.0014$ for glasses cooled through the 
transition at $10\ K.min^{-1}$ and quenched glasses \cite{Giridhar}.
\end{thebibliography}
\end{document}